\documentclass[reprint, amsmath, amssymb, aps,superscriptaddress]{revtex4-2}
\usepackage[utf8]{inputenc}
\usepackage{graphicx}
\usepackage[caption=false]{subfig}
\usepackage{xcolor}
\usepackage{hyperref}
\hypersetup{
colorlinks=true,       
linkcolor=cyan,          
citecolor=magenta,        
filecolor=magenta,      
urlcolor=cyan,           
runcolor=cyan
}
\newcommand{\ket}[1]{\ensuremath{\left\vert #1 \right\rangle}}
\newcommand{\bra}[1]{\ensuremath{\left\langle #1 \right\vert}}

\begin{document}
\title{Signatures of Open and Noisy Quantum Systems in Single-Qubit Quantum Annealing}
\author{Zachary Morrell}
\affiliation{Los Alamos National Laboratory, Los Alamos, NM 87545}
\author{Marc Vuffray}
\affiliation{Los Alamos National Laboratory, Los Alamos, NM 87545}
\author{Andrey Y. Lokhov}
\affiliation{Los Alamos National Laboratory, Los Alamos, NM 87545}
\author{Andreas B\"artschi}
\affiliation{Los Alamos National Laboratory, Los Alamos, NM 87545}
\author{Tameem Albash}
\affiliation{Department of Electrical \& Computer Engineering, University of New Mexico, Albuquerque, NM 87131, USA}
\affiliation{Department of Physics \& Astronomy, University of New Mexico, Albuquerque, NM 87131, USA}
\affiliation{Center for Quantum Information and Control (CQuIC), University of New Mexico, Albuquerque, NM 87131, USA}
\author{Carleton Coffrin}
\affiliation{Los Alamos National Laboratory, Los Alamos, NM 87545}

\begin{abstract}
We propose a quantum annealing protocol that more effectively probes the dynamics of a single qubit on D-Wave's quantum annealing hardware.  This protocol uses D-Wave's h-gain schedule functionality, which allows the rapid quenching of the longitudinal magnetic field at arbitrary points during the anneal. This features enables us to distinguish between open and closed system dynamics as well as the presence or absence of longitudinal magnetic field noise. We show that both thermal and magnetic field fluctuations are key sources of noise that need to be included in an open quantum system model to reproduce the output statistics of the hardware.
\end{abstract}

\maketitle

\section{Introduction}
Quantum annealing is an analog computing approach for preparing low-energy eigenstates of classical and quantum Hamiltonians \cite{qa_first_paper, qa_second_paper,Ray1989,San2002}. At the heart of the algorithm is the Adiabatic Theorem~\cite{born1928beweis,Kat1950,Jan2007}, which guarantees that a quantum system initially prepared in the ground state of a time-evolving Hamiltonian will remain with high probability in the instantaneous ground state at later times as long as the evolution satisfies an adiabatic condition. Quantum annealing exploits this property by slowly interpolating between a Hamiltonian for which the ground state is known and a target Hamiltonian that we wish to minimize \cite{Ray1989,qa_first_paper,San2002}. An advantage of quantum annealing over gate-based implementations of quantum computation is its minimal control requirements \cite{Chi2001,Sar2005,Abe2005,Rol2005}: generically, the qubits are annealed uniformly and slowly, which makes it in principle readily scalable to thousands of qubits. For this reason, quantum annealing remains a promising quantum optimization meta-heuristic in the NISQ era \cite{preskill_nisq}, with the hope that it may provide some advantage over classical algorithms. However, just like any other analog computing methods, quantum annealing is sensitive to hardware defects, limited controller accuracy \cite{Zhu2016,Alb2019} and other sorts of spurious effects and noise that are inherently present in any physical systems.

Today's largest and most mature quantum annealers are produced by D-Wave Systems with a qubit technology based on superconducting loops \cite{Joh2010,Ber2010,Har2010,Joh2011,Bun2014}. 
There is still an ongoing research effort to probe the effectiveness of the D-Wave hardware as an optimization tool \cite{Man2016,Man2018,pang2021potential,Kow2022}, as well as a Gibbs sampler \cite{amin2015searching, perdomo2016determination, raymond2016global, benedetti2016estimation, benedetti2017quantum, marshall2017thermalization, li2020limitations,ymca, john, nelson2022high}, which could prove useful in machine learning applications.
With the latest systems from D-Wave featuring thousands of qubits, it remains important to identify the noise sources and their effects on the output statistics in order to discover further use cases for the hardware.

In this paper, we test four different categories of dynamic models against D-Wave measurements and see if they have the power to reproduce the machine's behavior. These four categories aim at answering two different fundamental questions: how is the annealing dynamics affected by its environment (closed vs open systems), and how consistent the programmed Hamiltonian is over consecutive annealing runs (single vs mixture of random Hamiltonians). These models are effectively trying to reproduce the effect of noise over two different time scales, fast and slow with respect to the typical annealing time. Thermal fluctuations correspond to fast timescales since they account for changes in the state on the time scale of a single annealing run, while changes over multiple annealing runs -- being on larger time scales compared to a single anneal -- are modeled by a mixture of Hamiltonians with random longitudinal fields.

A lot of effort has been devoted to find signatures of thermal fluctuations in the D-Wave output statistics, which are typically modeled within an open system paradigm.
The advent of quantum annealing hardware has reignited development of open system descriptions for systems with time-dependent Hamiltonians (for recent advancements, see e.g. \cite{Smi2018,Moz2020,Che2022}). To model the effect of thermal fluctuations, we use the adiabatic master equation (AME)~\cite{Dav1978,ame} to describe the output statistics of the D-Wave hardware. The AME gives a description of dissipative dynamics under the assumption of a weak coupling to a finite temperature environment as well as slow evolution relative to the dynamical time-scales of the environment. This description has been successful at giving qualitative agreement for slow annealing processes on D-Wave processors \cite{PhysRevA.91.042314,PhysRevA.92.062328,ame_anneal_pause}.  The AME's success stems from the fact that the instantaneous steady state of the dynamics is the Gibbs state associated with the instantaneous Hamiltonian, hence it captures qualitatively the loss of population from the ground state due to thermal excitation when the ground state energy gap becomes smaller than the temperature, with the rate of population loss controlled by the system-bath coupling that can be tuned to fit experimental data as best as possible.  However, the AME's applicability is limited and is expected to breakdown near small energy gaps, where the weak-coupling assumption is violated.  Simulations using the polaron transformation \cite{Xu2016} (also called the non-interacting blip approximation \cite{Wei2012}) that allow investigations beyond the weak-coupling limit give the strongest quantitative agreement with the D-Wave hardware \cite{Boi2016}, but this approach tends to be limited to the lowest lying energy levels of the system.  The simplicity and interpretability of the AME makes it a convenient choice.

To model fluctuations in the programmed Hamiltonian, we use randomness in its parameters. Mixtures of random Hamiltonians with longitudinal field noise has been shown to play a major role in explaining the formation of the anomalous single qubit response to magnetic field changes~\cite{ymca, john} and of the effective spurious links~\cite{ymca, nelson2022high}.

One of the primary criticisms of the D-Wave hardware is that its output statistics can often be captured by simple classical models, such as the Spin-Vector Monte Carlo (SVMC) algorithm \cite{svmc,svmc_ibm} or Spin-vector Langevin dynamics \cite{PhysRevA.91.042314,PhysRevResearch.4.023104}. These methods provide descriptions of evolutions in the semi-classical energy landscape of the qubit system \cite{Kla1979}, which turns out to coincide well with the dynamics of superconducting flux qubits in the strong system-bath coupling limit \cite{Cro2016}. While these descriptions can quite often describe many qualitative features of the hardware, they do not necessarily reproduce all experimental results \cite{PhysRevA.92.062328}. Our aim in this work is not to make a claim that the output statistics can \emph{only} be described by a fully quantum open system description. Instead, we will show that a quantum description of the dynamics requires both thermal and magnetic field fluctuations to fully account for the system's behavior.

One main challenge in modeling the D-Wave hardware dynamics is that many experimental parameters are fixed and cannot be tuned by the user: the state preparation is predetermined and the system can only be measured at the end of the anneal in the computational basis. Therefore, there exists a substantial risk of over-fitting observations with different types of models when we consider large systems with multiple spins and complex interactions.  A key ingredient that we introduce to overcome data scarcity is what we call the $h$-stop protocol that currently can be performed only for systems without couplings. This modification of the annealing protocol effectively enables us to get information about the state of the system during the dynamics. 
Leveraging this new protocol, we will demonstrate that both \emph{thermal fluctuations} and \emph{longitudinal field noise} are necessary ingredients to explain the dynamic properties measured in the D-Wave. In particular, we show that the $h$-stop protocol dynamics cannot be reproduced with thermal fluctuations alone even though it can be approximated with closed system dynamics and longitudinal field noise at small annealing time and input magnetic field. This highlights the importance of slower fluctuations in the D-Wave quantum annealing dynamics.

\section{Background and Dynamic Models}

\begin{figure} 
\centering
\includegraphics[width=\linewidth]{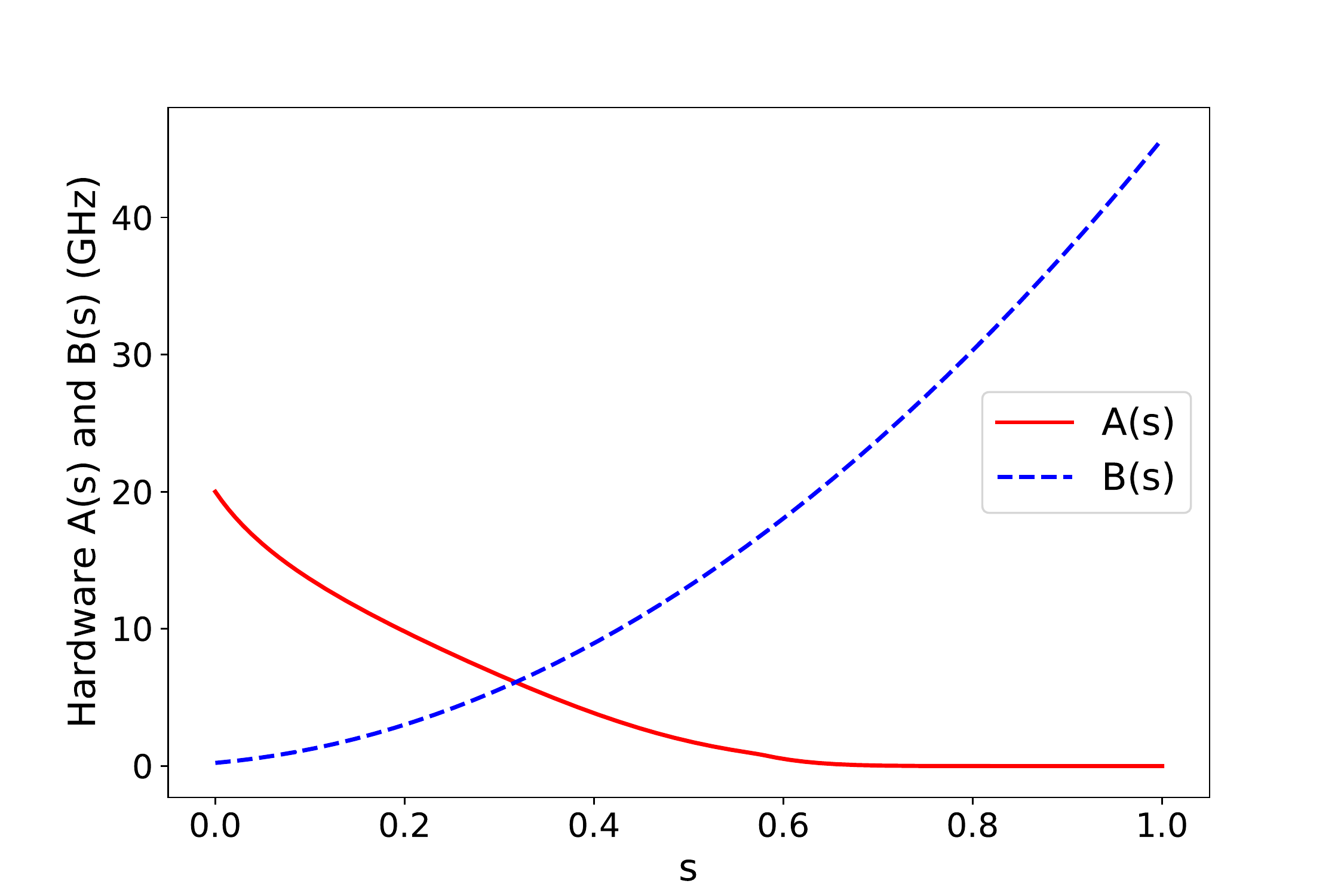}
\caption{The D-Wave default annealing schedule for the DW\_2000Q\_LANL system. The red line and the blue dashed line represent the annealing schedules $A(s)$ and $B(s)$ in Eqt.~\eqref{eq:Dwave_hamiltonian}, respectively. The unit of energy of the annealing schedule is expressed in Hertz after setting $\hbar=1$.}
    \label{fig:standard_anneal}
\end{figure}

The Hamiltonian realized in the D-Wave quantum annealers is the transverse field Ising Model
\begin{equation}
    H(s) = -A(s)\sum_i{\xi_i \sigma_i^x} + B(s)\left(\sum_i{h_i \sigma_i^z} + \sum_{i < j} J_{ij} \sigma_i^z\sigma_j^z \right)\label{eq:Dwave_hamiltonian}
\end{equation}
where the annealing schedule is defined by functions $A(s)$ and $B(s)$, with normalized time parameter $s = t/\tau$, where $t \in [0,\tau]$ is the current time in the anneal and $\tau$ is the total anneal time. The programmed transverse field strength and the programmed longitudinal field strength on qubit $i$ are given by $\xi_i$, and $h_i$ respectively. The coupling strength between qubits $i$ and $j$ is given by $J_{ij}$.  The user is only able to modify the programmed longitudinal field strengths and coupling strengths.  The Pauli operators on each qubit are given by $\sigma^x_i$ and $\sigma^z_i$, and measurements are performed in the computational basis, such that $\ket{0} = \ket{\uparrow}$ and $\ket{1}= \ket{\downarrow}$. The state of the system is described by its density matrix $\rho(s)$ which is initially prepared in the ground state of the Hamiltonian at $s=0$, i.e., $\rho(0) = \ket{+}\bra{+}$, where $\ket{+} = \bigotimes_{i} \frac{1}{\sqrt{2}} \left(\ket{\uparrow} + \ket{\downarrow}\right)$.
The annealing protocol slowly interpolates between a transverse field Hamiltonian and the target Ising Hamiltonian on the longitudinal component, i.e., $A(0) \gg B(0)$ and $A(1) \ll B(1)$, see Fig~\ref{fig:standard_anneal}. It is customary to measure energy in units of $\hbar$ (or equivalently setting $\hbar = 1$), with the convention that fields and couplings are dimensionless and the annealing is expressed in Hertz.

For our experiments, we made use of the DW 2000Q Hardware from D-Wave Systems which was maintained by Los Alamos National Laboratory.  On the 2000Q chip, D-Wave uses superconducting flux qubits to implement a quantum annealer with a so-called Chimera graph connection topology \cite{Cho2008,Bun2014}. In the present work we focus our attention on single qubits; the topology plays no role in our study as we set all coupling strengths to $J_{ij} = 0$.

In what follows, we go through the different dynamic models that we consider to encode the time evolution of the density matrix.

\subsection{Closed Quantum Systems}
A closed quantum system simulation assumes that the interactions between the environment and the system are negligible. Therefore, the evolution of the density matrix is described by the von Neumann equation
\begin{equation}
    \frac{d}{ds} \rho(s) = -i\tau[H(s),\rho(s)],
    \label{eq:Neumann}
\end{equation}
where we have set $\hbar = 1$. We solve the von Neumann equation numerically using a 4th order integration method based on the Magnus expansion \cite{BLANES2009151}. Eq.~\eqref{eq:Neumann} is the simplest way to represent the evolution of a quantum system.
However, interactions with the environment are often unavoidable, so it is critical to understand what behavior can be induced by such interactions.
This leads to the challenge of simulating an open quantum system.

\subsection{Open Quantum Systems}

We choose to model the open system dynamics using the AME \cite{ame}, which is a master equation in Lindblad form \cite{Lin1976} given by:
\begin{eqnarray}
    \frac{1}{\tau}\frac{d}{ds} \rho(s) = &&\   -i\left[H(s),\rho(s)\right] \nonumber\\
    &&\ + \sum_i \sum_{\omega} \gamma(\omega) \left[ L_{\omega i}(s)\rho(s)L_{\omega i}^{\dagger}(s) \phantom{\frac{1}{2}} \right. \nonumber\\
    &&\ \left. - \frac{1}{2} \left\{ L_{\omega i}^{\dagger}(s)L_{\omega i}(s), \rho(s) \right\} \right] \label{eq:master}
\end{eqnarray}
where $\gamma(\omega)$ encodes properties of the bath and satisfies the KMS condition $\gamma(-\omega) = e^{-\beta \omega} \gamma(\omega)$ \cite{Kub1957,Mar1959,Haa1967},  the sum over $\omega$ is the sum over all Bohr frequencies (differences of all possible energy eigenvalues of $H(s)$), and the sum over $i$ is over all system-bath interaction terms. We assume that each qubit interacts with an independent yet identical Ohmic heat bath of harmonic oscillators, such that
\begin{equation}
    \gamma(\omega) = 2\pi g^2 \frac{\omega e^{-|\omega|/\omega_c}}{1 - e^{-\beta \omega}}
\end{equation}
where $\omega_c$ is the cutoff frequency, and the Lindblad operators are then given by
\begin{equation}
    L_{\omega i} = \sum_{a,b}  \delta_{\omega,E_b(s) - E_a(s)} \langle E_a(s)| \sigma_i^z | E_b(s) \rangle |E_a(s)\rangle \langle E_b(s)|
\end{equation}
where we have assumed a $\sigma^z$ system-bath interaction for each qubit and ${|E_{a}(s)\rangle}$ are the instantaneous eigenstates of the Hamiltonian with eigenvalues $E_a(s)$.

\subsection{Longitudinal Field Noise}
The density matrix $\rho(s)$ can only be observed indirectly through projective measurements. In order to obtain good quality estimates of the outcome probabilities, it is necessary to repeat the same experiment a large number of times. However, interactions in the system can slowly change between runs due to limitations on the controller accuracy or slow evolving exogenous sources of noise. We model this process using a mixture of random Hamiltonians, where fluctuations take place on the longitudinal field. More precisely, the random Hamiltonian describing a single qubit is modeled through the following expression,
\begin{equation} \label{eqt:fluctH}
    H(s) = - A(s) \xi \sigma^x + B(s) h \sigma^z  + B(1) \Delta z \sigma^z, 
\end{equation}
where $\Delta z \sim \mathcal{N}(\mu_{\Delta z}, \sigma_{\Delta z})$ represents a longitudinal field that is constant and independent of time during a single anneal but takes a random value for different anneals.
The random contribution is rescaled by $B(1)$ allowing for the mean and standard deviations of the distribution to be interpreted by their scale relative to the maximum programmed longitudinal field. 
The simulation of the mixture of random Hamiltonian is realized by running either the von Neumann Eq.~\eqref{eq:Neumann} or the AME~\eqref{eq:master} 1000 times with $\Delta z$ being resampled for each simulation. The final density matrix of the mixture is the average of the density matrices that we obtained for different values of $\Delta z$.

\section{Experiments and Results}
We test the predictive power of our models that describe a closed or an open quantum system with and without longitudinal field noise. For simplicity, we will refer to these models as closed/open and noisy/noiseless models. Our experiments consist of collecting the output probability distribution of a single qubit system for different value of the input parameters then comparing the data to the predictions of our different models where the parameters of the models are fitted to best reproduce the experimental results. More details on the fitting procedure can be found in Appendix~\ref{app:model_fitting}. We consider two different data collection and annealing protocols: the $h$-sweep protocol and the $h$-stop protocol. All of the data was obtained from the now decommissioned DW\_2000Q\_LANL system.

\subsection{$h$-sweep Protocol}
\begin{figure*}
    \centering
        \subfloat[$\tau = 1 \mu s$]{
        \includegraphics[width=0.45\linewidth]{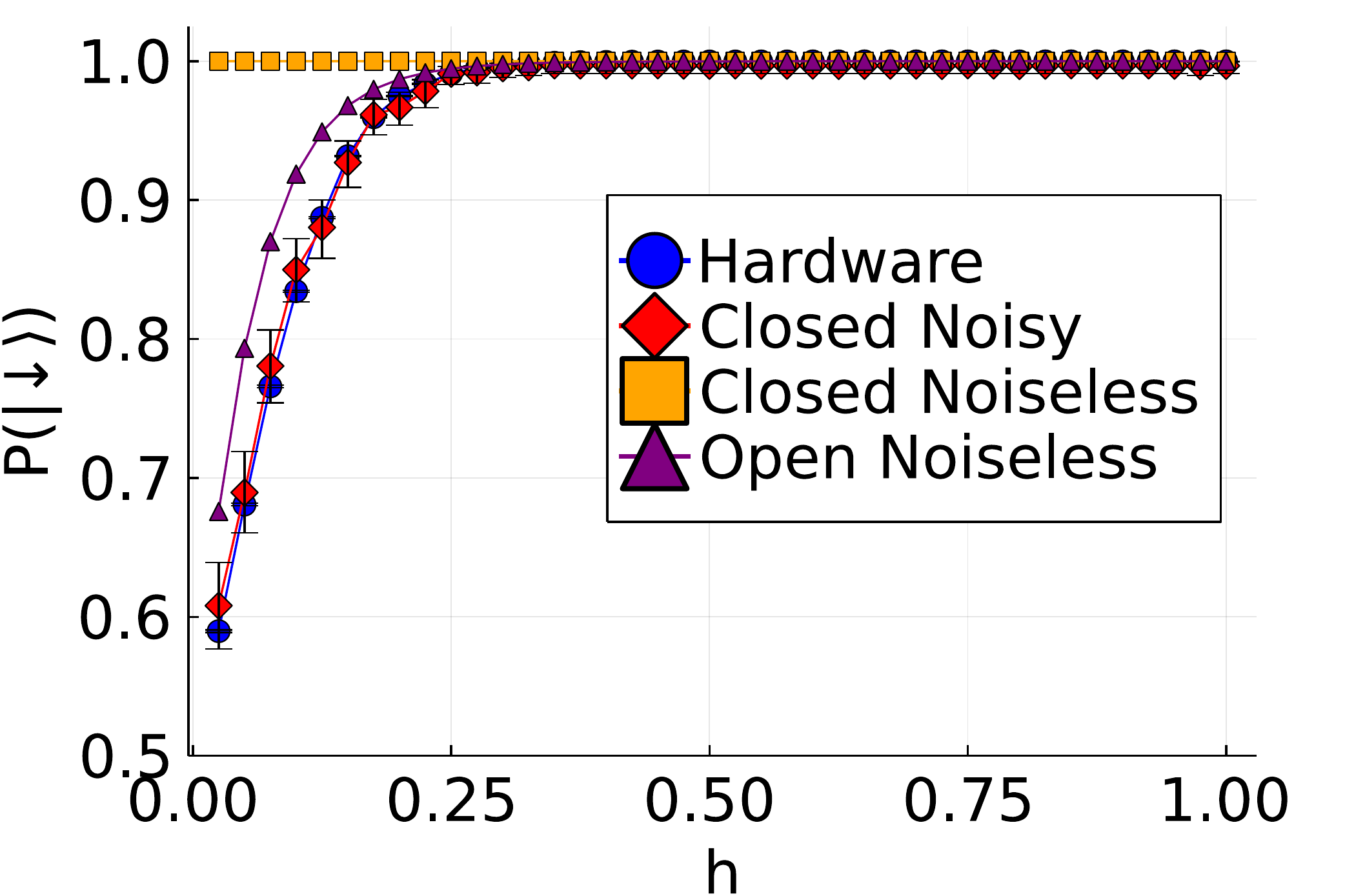}
        \label{fig:magnus_hsweep_sub_1}
        }
        \subfloat[$\tau = 125 \mu s$]{
        \includegraphics[width=0.45\linewidth]{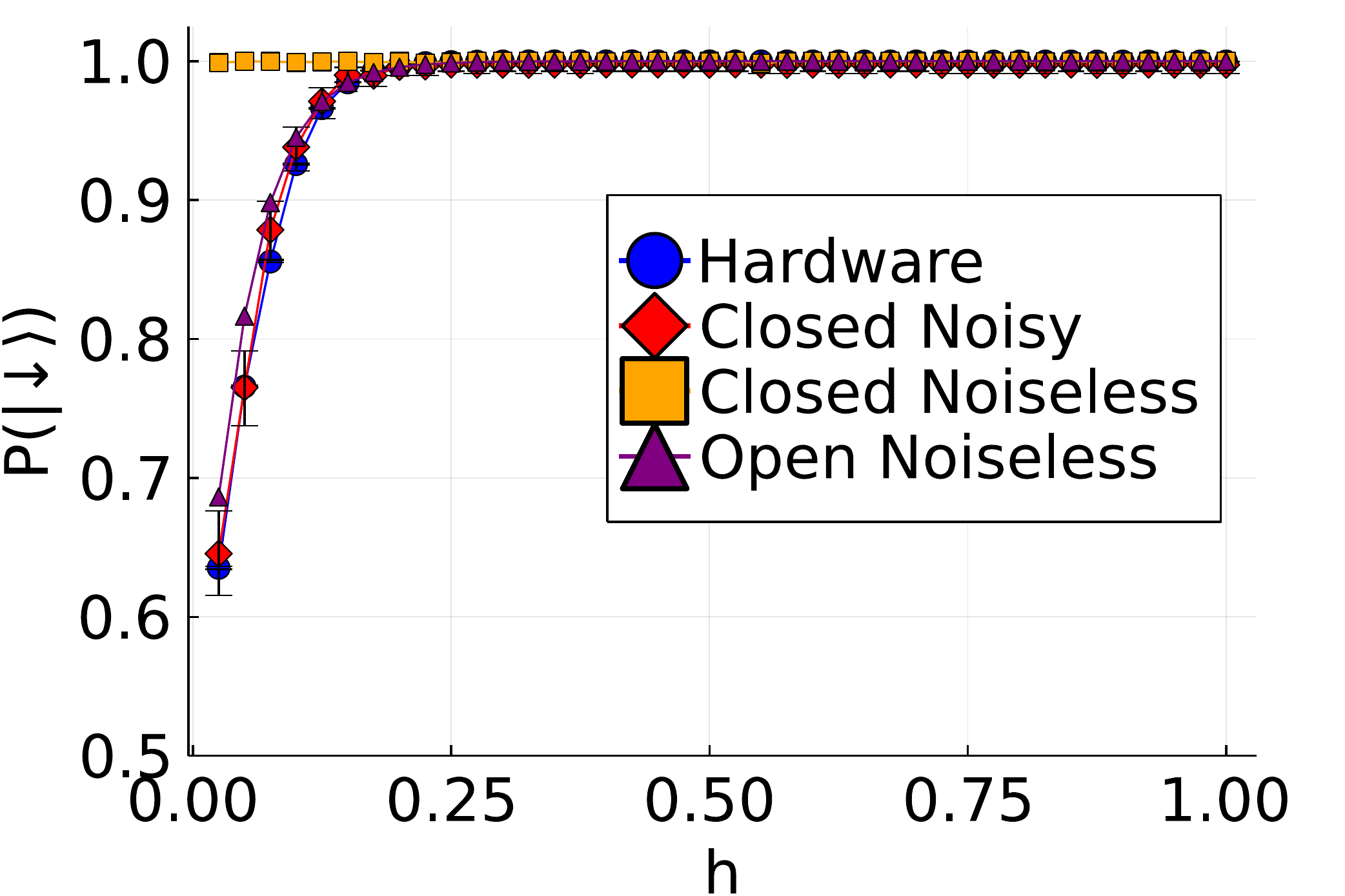}
        \label{fig:magnus_hsweep_sub_2}
    }
    \caption{Measurement probabilities $P(\ket{\downarrow})$ at the end of the anneal with respect to input magnetic field $h$. The blue line represents the data sampled from the LANL\_DW2000Q hardware with one million shots, the red line represents the simulation of the closed noisy system, the green line represents the results of the closed noiseless system, i.e. $\Delta z = 0$, and the purple line shows the simulation of the open noiseless system.  This figure does not include any open noisy system simulations because the closed noisy systems simulations already fit the data, implying open noisy system fits would be degenerate.}
    \label{fig:magnus_hsweep_examples}
\end{figure*}

The default annealing protocol of D-Wave's API uses the annealing schedule functions shown in Fig.~\ref{fig:standard_anneal}. 
We question what sort of noise signatures can be observed with this default annealing schedule and whether these signatures can be used to differentiate between an open and a closed quantum system's evolution.
For a single spin, the two parameters that can be modified are the annealing time $\tau$ and the longitudinal field $h$. For different annealing time, we sweep over several $h$ values to see how the output probabilities change, as was done in \cite{john}. We refer to this type of experiment as the $h$-sweep protocol.
For our $h$-sweep experiments, we took one million samples for each $h$ value, sweeping over $h = \{0.025, 0.050, \dots, 1.0\}$. 

The results of the $h$-sweep protocol and the best fit for closed systems with and without random longitudinal field noise are shown in Fig.~\ref{fig:magnus_hsweep_examples}.
The hardware exhibits a drop in the probability of measuring $\ket{\downarrow}$ at small $h$, which the closed-noiseless simulations do not exhibit.  For the closed-noiseless simulations, the evolution is adiabatic even for the lowest $h$ and $\tau$ values used, so the probability is effectively 1. However, the closed-noisy and open-noiseless simulations reproduce the drop in probability of the hardware for distinctly different reasons. In the open-noiseless case, thermal excitations out of the ground state are less suppressed at small $h$ leading to a loss of population to the $\ket{\uparrow}$ state.  For the closed-noisy case, the noise can more readily change the direction of the longitudinal field at small $h$, leading to a large fraction of the runs resulting in the $\ket{\uparrow}$ state.
This experiment demonstrates that the single-qubit dynamics cannot be reproduced by a simple closed system model, particularly for small values of the magnetic field. However, interestingly, a closed-and-noisy model cannot be distinguished from an open noiseless model with this type of experiment. Thermal fluctuations or random mixture of longitudinal fields induce a similar type of behavior in response to a change in the input magnetic field.

\subsection{$h$-stop Protocol}

The $h$-sweep protocol demonstrates the limitations of the default annealing schedule in discriminating between different dynamical models, and it is necessary to implement a richer class of protocols. One possible approach is to use D-Wave's annealing-schedule controls to hold $A(s)$ and $B(s)$ constant for a certain period of time to allow for thermalization effects to take place, as is done by Ref.~ \cite{ame_anneal_pause}. Unfortunately, a protocol like this generally requires the use of more qubits in order to glean useful information, which can lead to model over-fitting and becomes challenging to simulate efficiently when performing thousands of noise realizations. 

Another approach is to use an annealing schedule with a quench at various $s$ values throughout the anneal to approximate instantaneous measurements, as described in Ref.~\cite{dwave_quench_docs}. This procedure hopes that the system will not have the time to respond to the rapid changes in the annealing rate. Unfortunately, the current quench rates on the hardware are not short enough for single qubit quenches, since the system still tends to reach the final ground state with high probability.

We will show that it is possible to obtain interesting measurement statistics by making use of D-Wave's h-gain schedule functionality \cite{dwave_h_gain}.
D-Wave's h-gain schedule control works by introducing a user defined function, $k(s)$ on the $h$ parameter in the Hamiltonian. In the case of the single qubit model with longitudinal field noise, the Hamiltonian from Eq.~\eqref{eq:Dwave_hamiltonian} is modified to,
\begin{equation}
    H(s) = - A(s) \xi \sigma^x + B(s) k(s) h \sigma^z  + B(1) \Delta z \sigma^z.
\end{equation}
The idea behind this modified scheduled is to obtain instantaneous snapshots of the hardware state by manipulating the $k(s)$ function, in particular, by setting $k(s)$ to zero at a certain moment during the anneal. Note, however, that this procedure does not fully suspend the dynamics since the transverse field remains untouched, so our measurements are not equivalent to an instantaneous measurement.

\begin{figure}
    \centering
    \includegraphics[width=\linewidth]{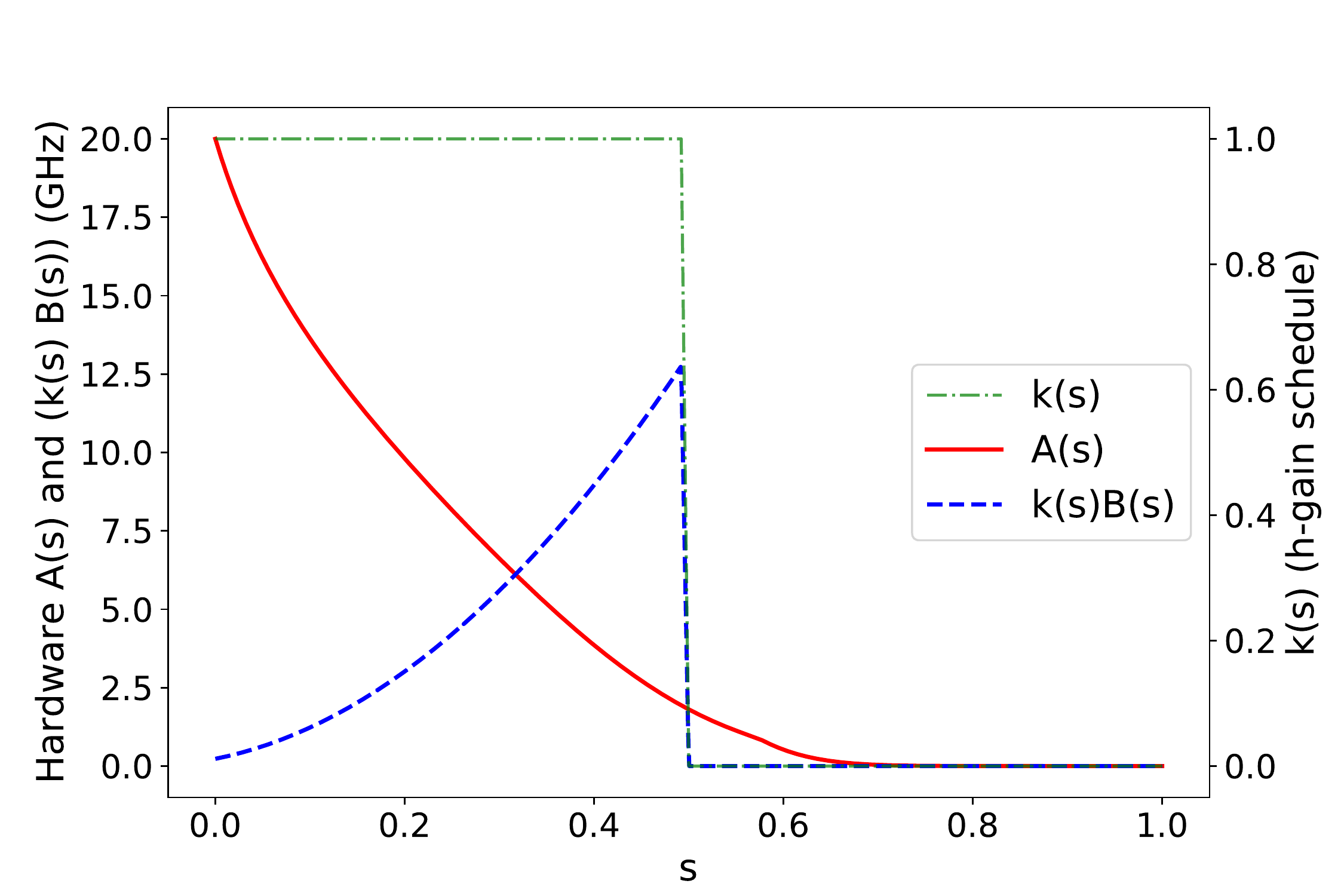}
    \caption{Annealing schedule with the $h$-stop protocol.  In this case, the h-gain stop finishes at $s_{\text{stop}} = 0.5$.  It should be emphasized that the annealing schedule effects shown here only apply to systems where there are no coupling terms $J_{ij}$, since there is no coupling-gain schedule control on the D-Wave hardware at the time of this writing.}
    \label{fig:hgk_anneal}
\end{figure}

The protocol we propose is rather simple: anneal with the standard annealing schedule (no pauses or quenches), and use the h-gain schedule parameter to eliminate the longitudinal field at various $s$ values throughout the anneal.  We will call this the $h$-stop protocol, since the user ``stops" the $h$ parameter at various points in the anneal.  This protocol is a very effective way to suppress the main driving term of the dynamics because the $k(s)$ value can be reduced to 0 as quickly as 2 nanoseconds after it is at full strength - far faster than the $0.5 \mu s$ needed to perform a quench~\cite{dwave_quench_docs}.  In our experiments we chose to use an 8 nanosecond stop time to avoid potential issues from operating at the edges of the hardware's capability.  An example of this annealing schedule can be seen in Fig.~\ref{fig:hgk_anneal}.

In our experiments, we collected data for h-stops at points $s_{\text{stop}} \in \{ 0.02, 0.04, \dots, 0.98 \}$, with one million data points per $s_{\text{stop}}$ and an $h$-stop time of 8 nanoseconds.  We sampled in this manner for all input combinations of $h = \{ 0.025, 0.05, 0.125, 0.25, 0.5 \}$ and anneal times $\tau = \{ 1, 5, 10, 25, 50, 125 \}$ microseconds.

\begin{figure*}
    \centering
    \subfloat[$\tau =  1 \mu\text{s}$, $h = 0.025$]{
        \centering
        \includegraphics[width=.32\linewidth]{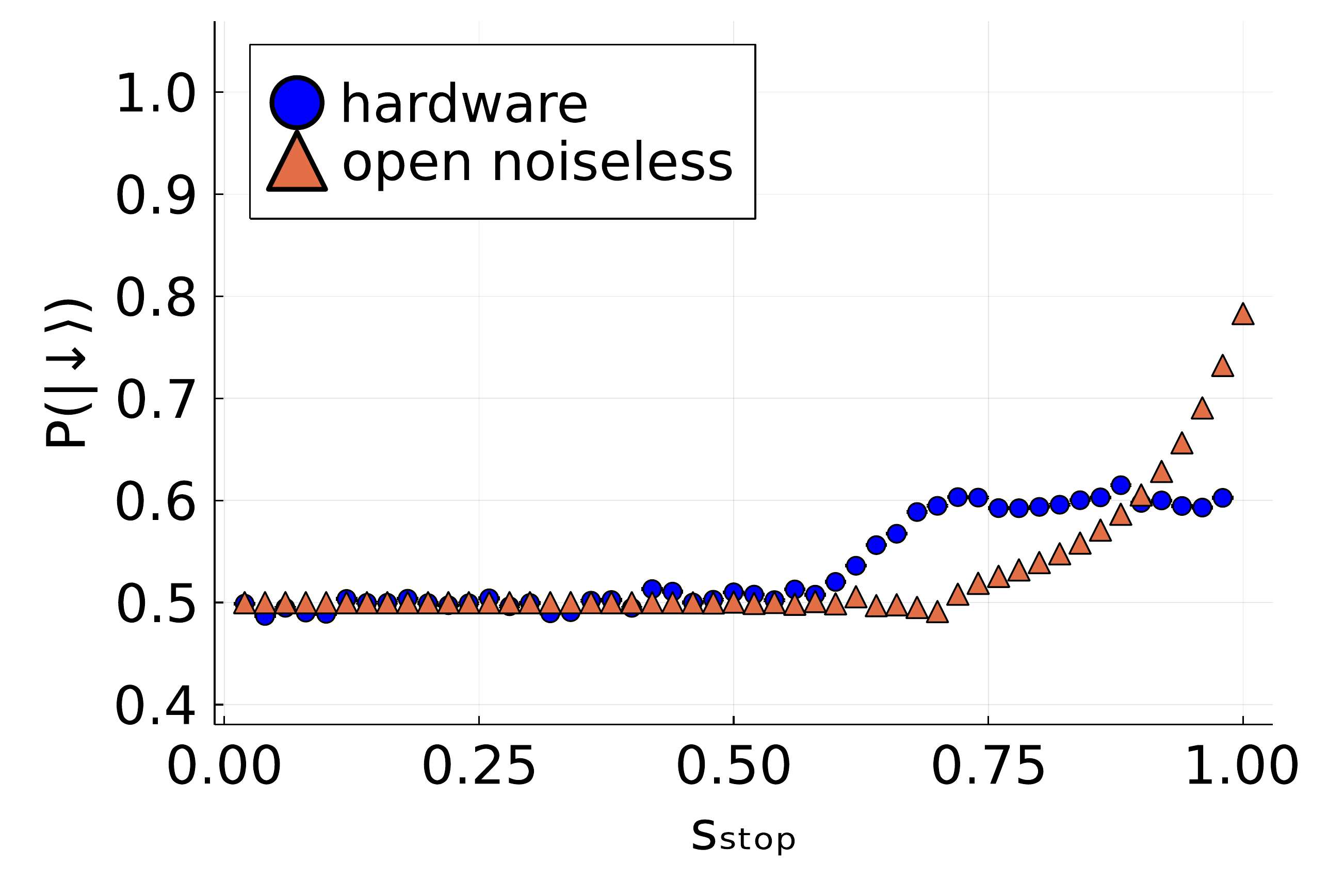}
        \label{fig:open_noiseless_ssweep_sub_1}
    }
    \subfloat[$\tau =  10 \mu\text{s}$, $h = 0.125$]{
        \centering
        \includegraphics[width=.32\linewidth]{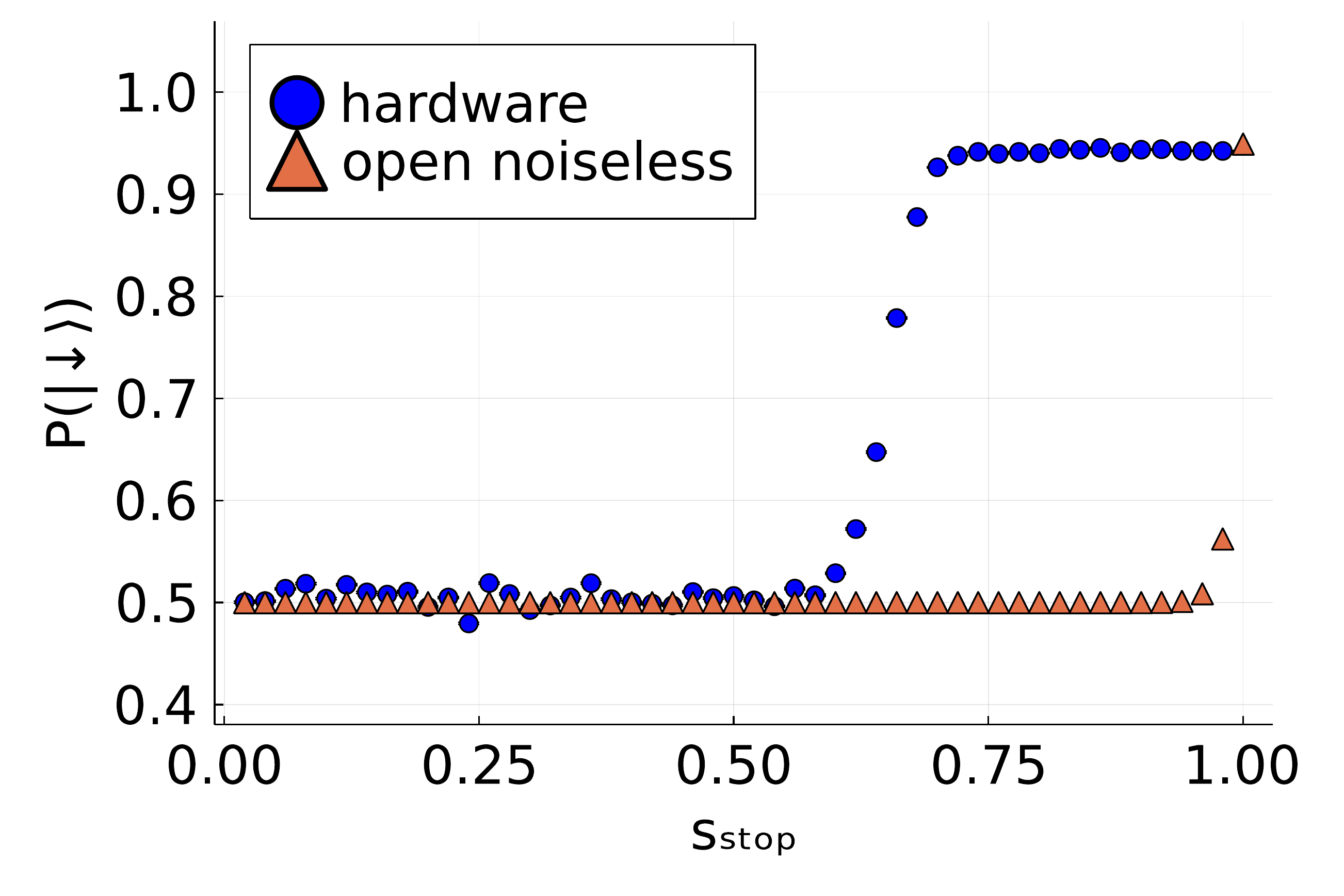}
        \label{fig:open_noiseless_ssweep_sub_2}
    }
    \subfloat[$\tau =  125 \mu\text{s}$, $h = 0.5$]{
        \centering
        \includegraphics[width=.32\linewidth]{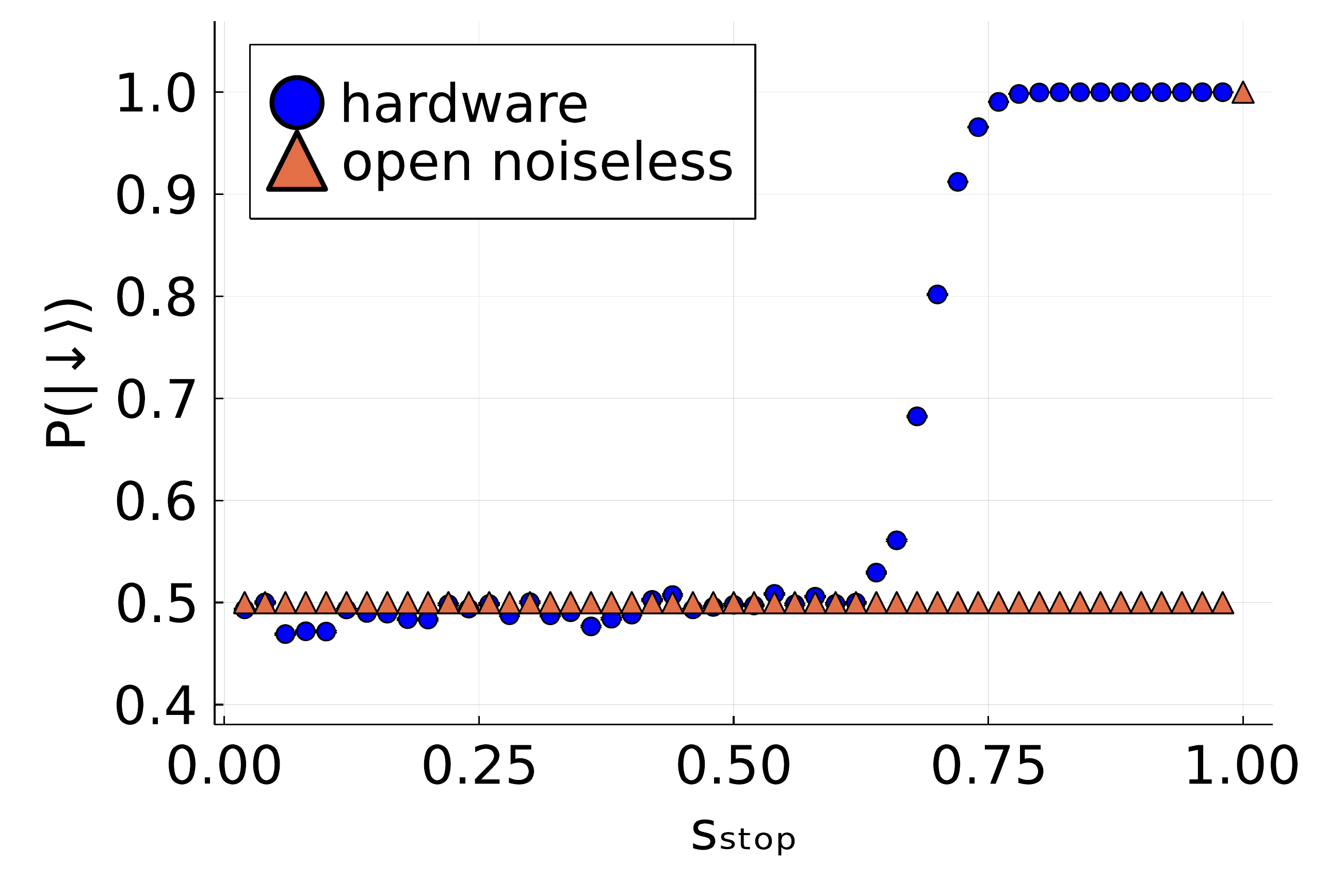}
        \label{fig:open_noiseless_ssweep_sub_3}
    }
    \\
     \centering
   \subfloat[$\tau = 1 \mu\text{s}$, $h = 0.025$]{
        \centering
        \includegraphics[width=.32\linewidth]{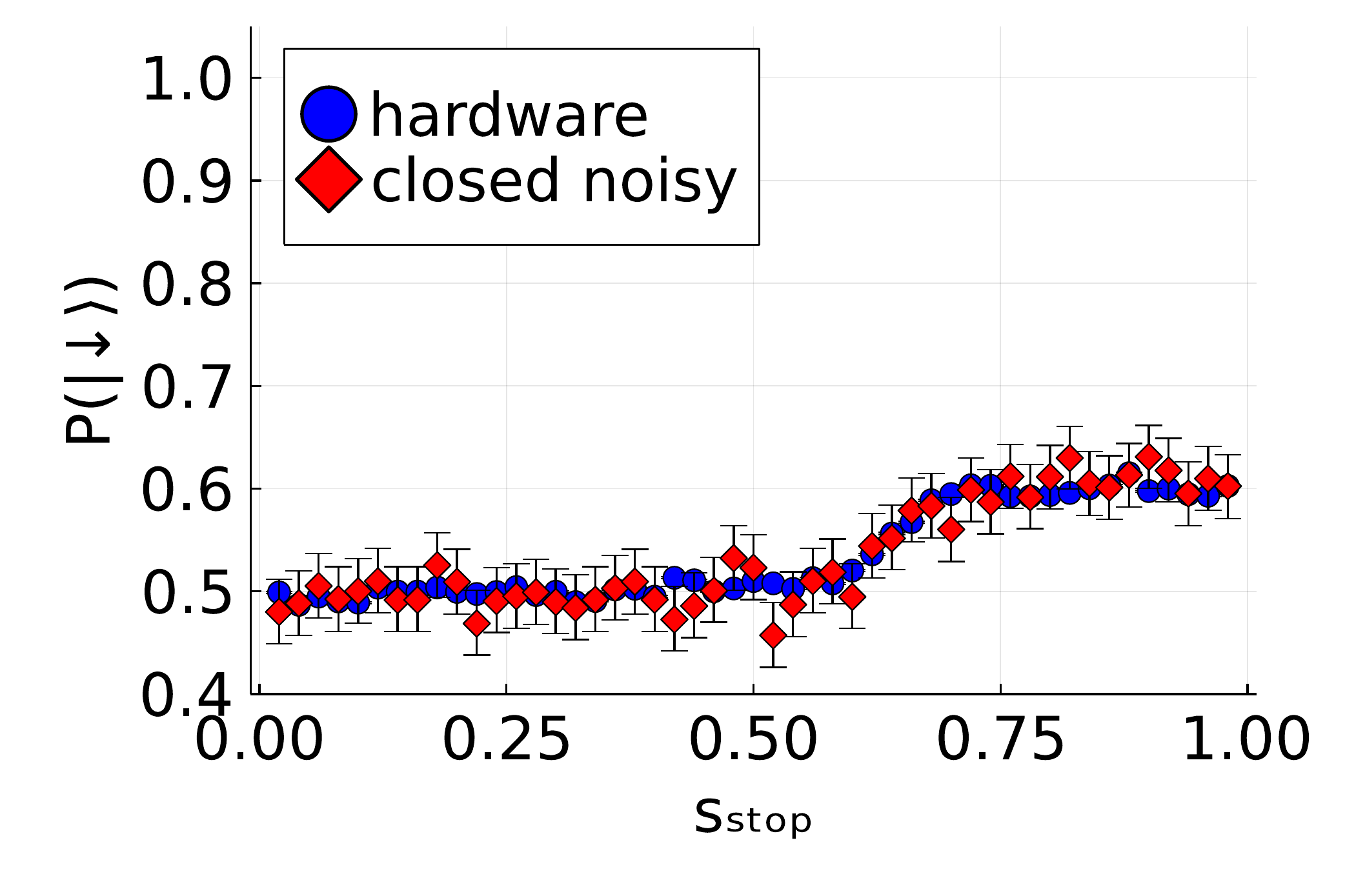}
        \label{fig:closed_ssweep_sub_1}
    }
    \subfloat[$\tau =  10 \mu\text{s}$, $h = 0.125$]{
        \centering
        \includegraphics[width=.32\linewidth]{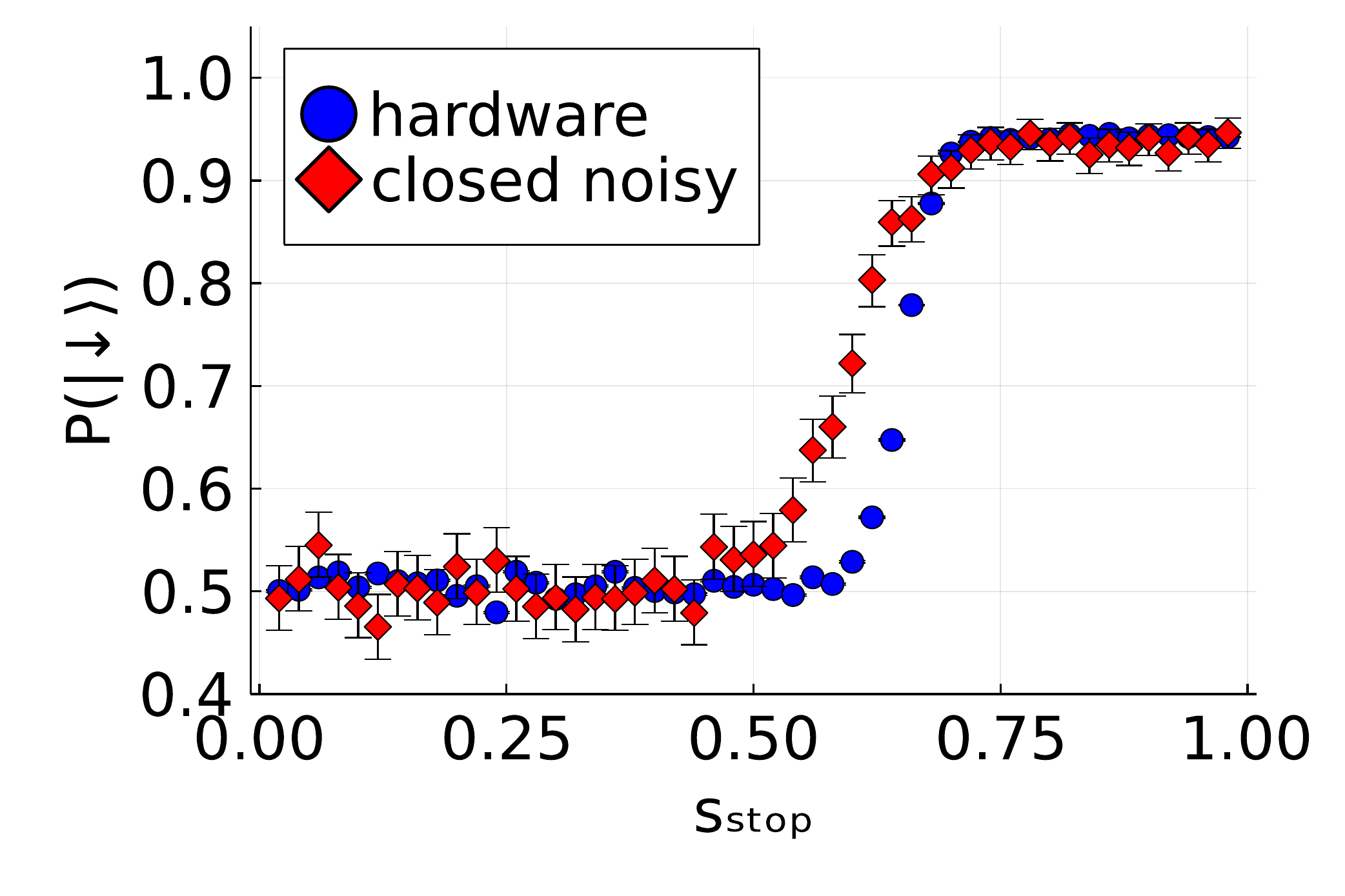}
        \label{fig:closed_ssweep_sub_2}
    }
    \subfloat[$\tau =  125 \mu\text{s}$, $h = 0.5$]{
        \centering
        \includegraphics[width=.32\linewidth]{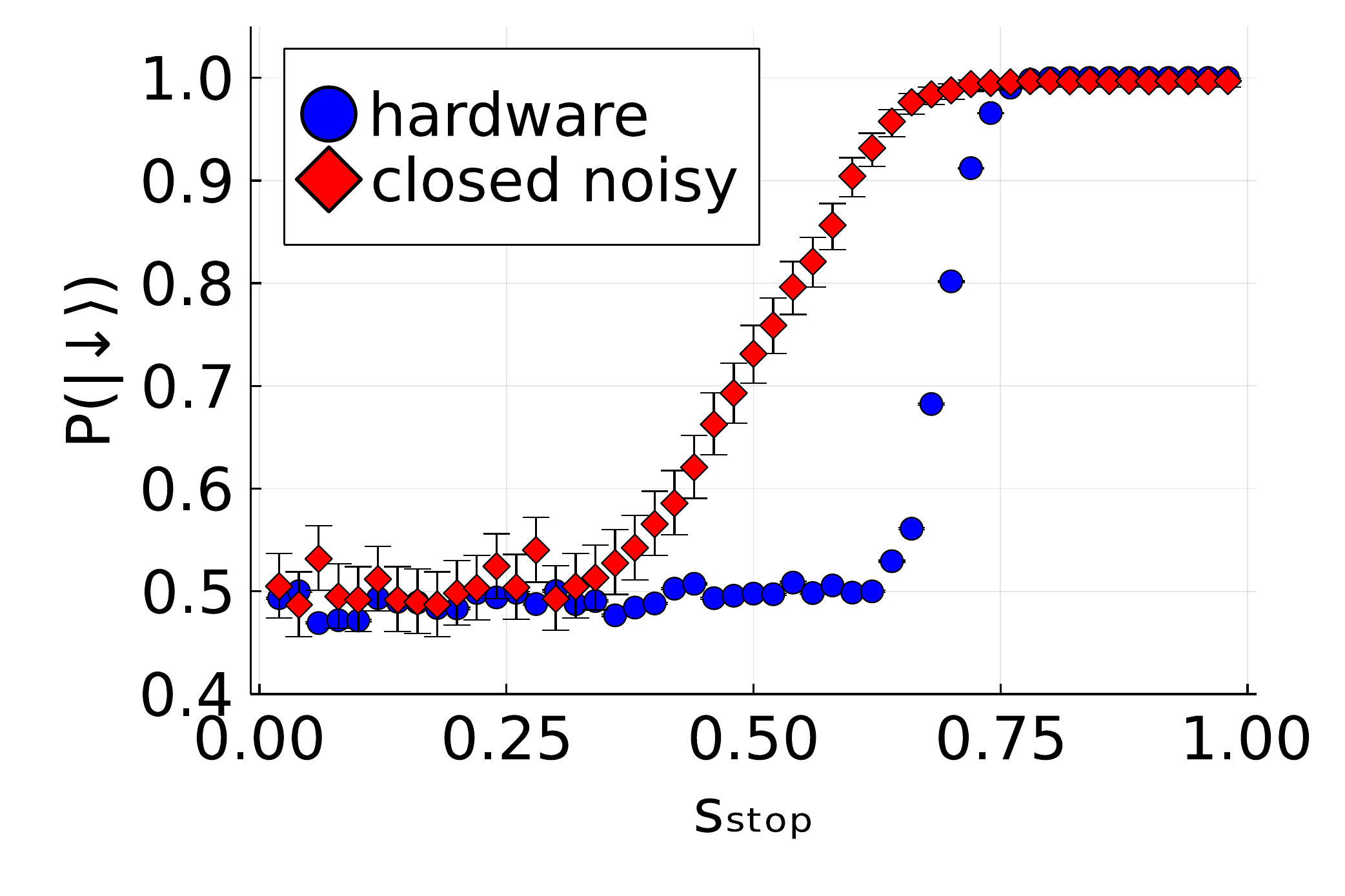}
        \label{fig:closed_ssweep_sub_3}
    }
    \\
    \centering
    \subfloat[$\tau =  1\mu\text{s}$, $h = 0.025$]{
        \centering
        \includegraphics[width=.32\linewidth]{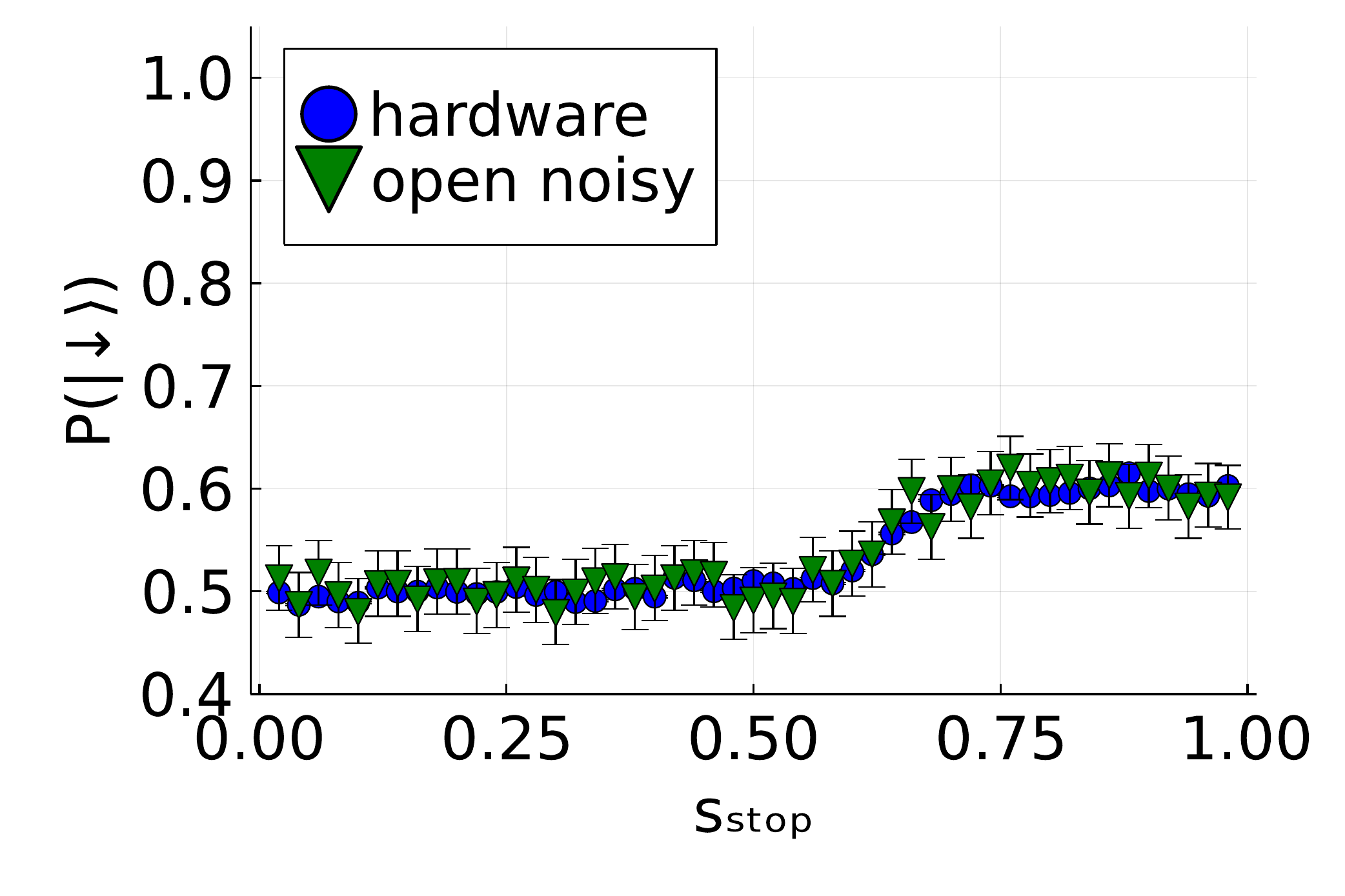}
        \label{fig:open_ssweep_sub_1}
    }
    \subfloat[$\tau =  10 \mu\text{s}$, $h = 0.125$]{
        \centering
        \includegraphics[width=.32\linewidth]{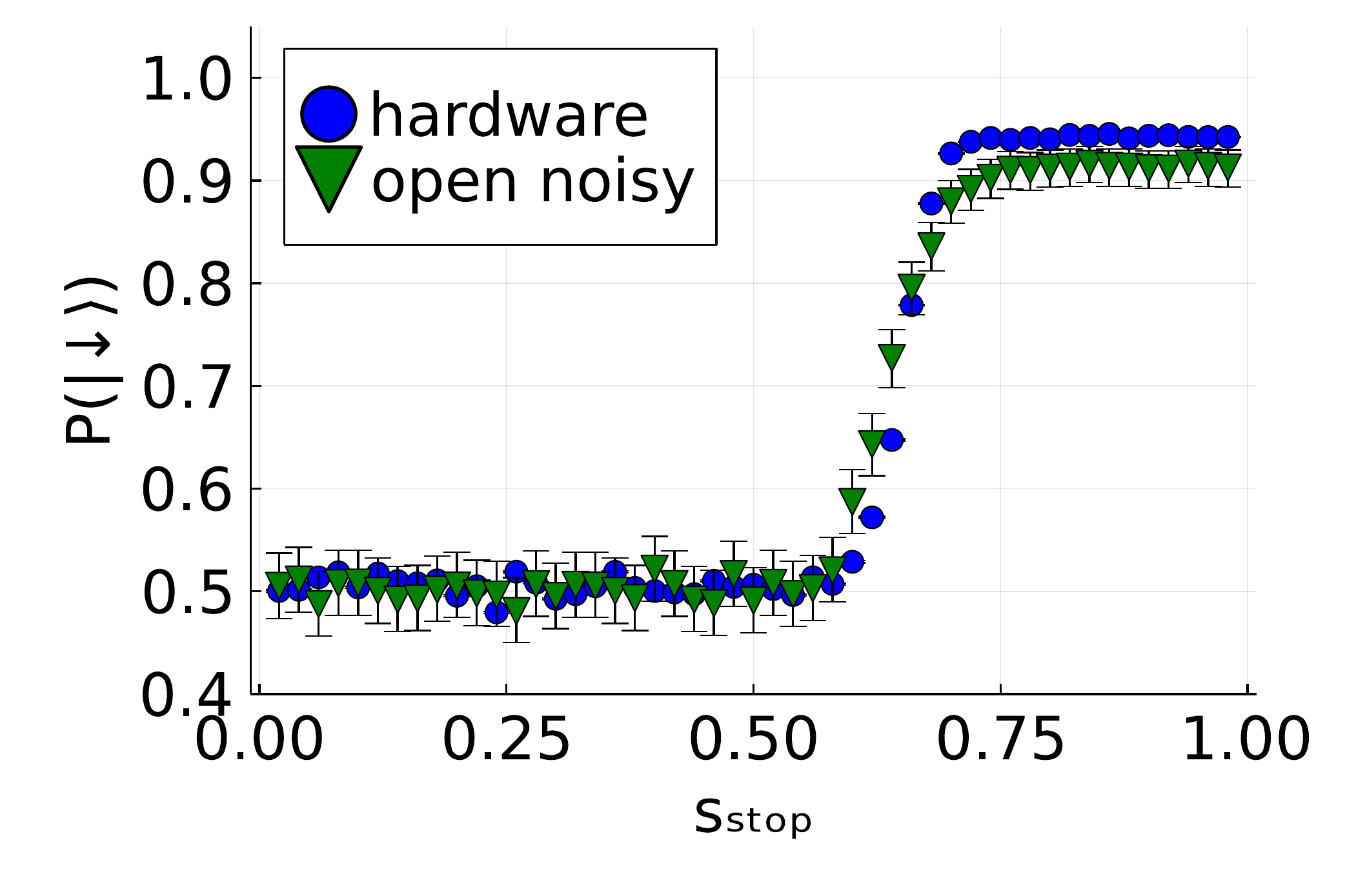}
        \label{fig:open_ssweep_sub_2}
    }
    \subfloat[$\tau =  125 \mu\text{s}$, $h = 0.5$]{
        \centering
        \includegraphics[width=.32\linewidth]{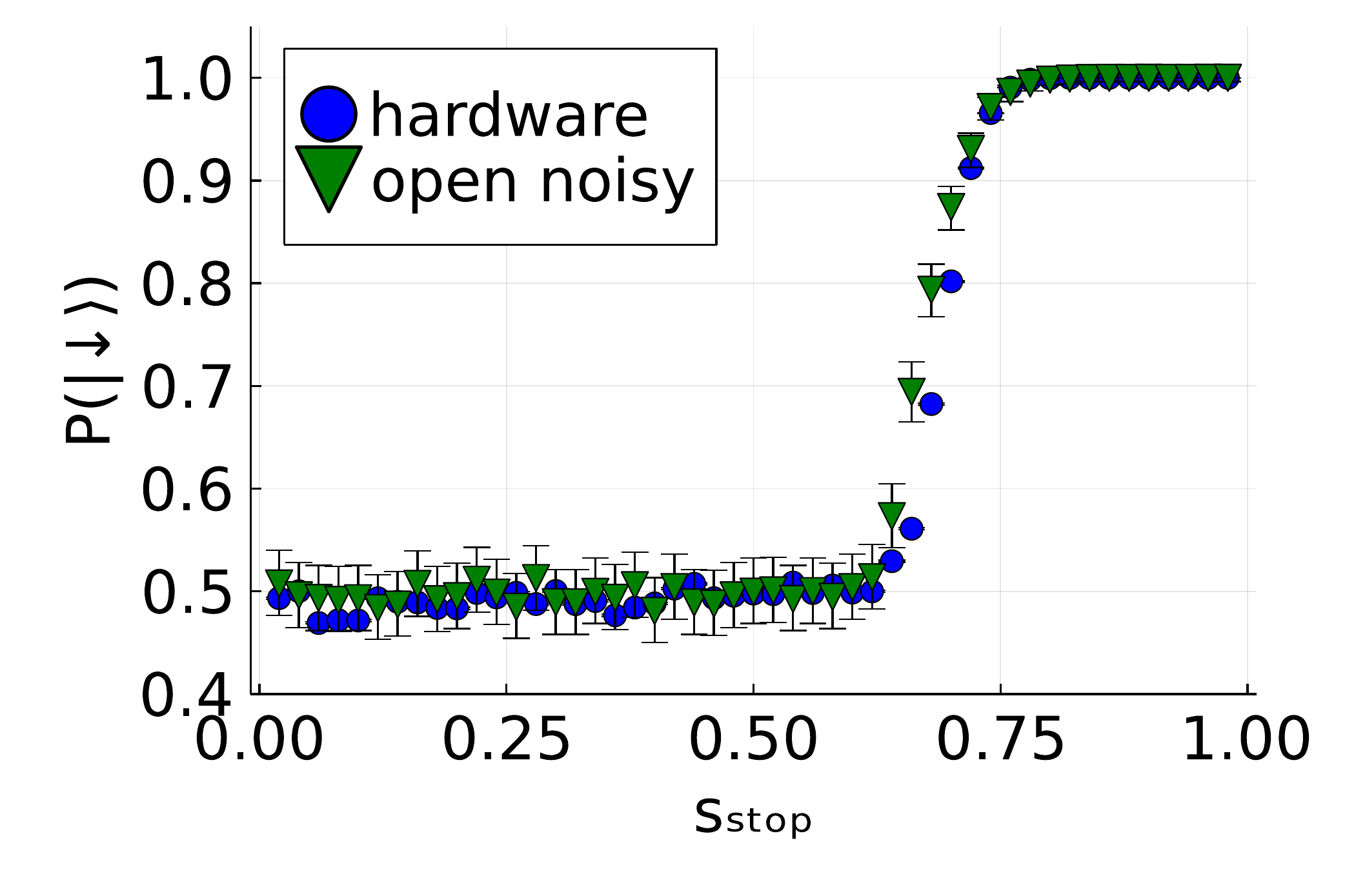}
        \label{fig:open_ssweep_sub_3}
    }
    \caption{ 
    Measurement probabilities $P(\ket{\downarrow})$ with respect to $s_\text{stop}$. Each column represent a different values of the annealing time $\tau$ and input magnetic field $h$. From left to right column $(\tau,h)$ is $(1\mu s,0.025)$, $(10\mu s,0.125)$ and $(125\mu s,0.5)$. Hardware data are shown in blue and are repeated over rows. Each row is associate with one particular model whose best fit is shown in red. From top to bottom we have the open noiseless model, the closed noisy model and the open noisy model.
    }
    \label{fig:ssweep}
\end{figure*}

Hardware results for the $h$-stop protocol and the model fits for the open noiseless, closed noisy, and open noisy models are shown in Fig~\ref{fig:ssweep} for some values of $\tau$ and $h$. It is important to note that the model parameters are fitted independently in each of the three different scenario. While sweeping over $s_\text{stop}$ values, we distinguish three main stages in the behavior of the system: the initial plateau, the transient, and the saturation. The initial plateau is located at around $P(\ket{\downarrow}) = 0.5$. This plateau at small $s_{\text{stop}}$ values can be expected. The system is prepared approximately in the state $\frac{1}{\sqrt{2}}\left(\ket{\uparrow} + \ket{\downarrow}\right)$, and for $s < s_{\text{stop}}$ the Hamiltonian component along $x$ dominates over that along $z$ so that the dynamics does not change the state appreciably.  When the local field is stopped, the dynamics is still not significantly changed since the transverse field was so much larger.

The transient phase takes the form of a sigmoid function and it has two important features - the $s_{\text{stop}}$ where the transition begins, and the slope of the transition. 
Finally, the saturation phase displays a relatively constant behavior with a saturation point that tends to $P(\ket{\downarrow}) = 1$ as $\tau$ or $h$ increases. As we will see from our model-fitting shortly, these features are a consequence of the fluctuating local field.

The worst agreement with the hardware comes from the open noiseless model in Fig~\ref{fig:ssweep}. It is able to saturate at the correct observed value when $s_\text{stop}=1$, which explains why it is able to reproduce the system's behavior in the $h$-sweep protocol in Fig~\ref{fig:hgk_anneal}. However, it does so only at the very end of the anneal and therefore fails to display a transient phase in the middle of the experiment. This behavior is straightforward to understand.  When the local field is stopped, the dissipative dynamics drives the system towards the thermal state of the transverse field Hamiltonian.  The longer the anneal, the more time the system has to thermalize after the transverse field is turned off, but this time is reduced as $s_{\text{stop}}$ is pushed towards 1.  The absence of a plateau in this model and its presence in the other noisy model indicates how crucial a fluctuating local field is in determining the output statistics.

The closed-noisy model in Fig~\ref{fig:ssweep} is able to reproduce the whole dynamics for low annealing time and low input $h$ value. 
In particular, it is able to reproduce the plateau at large $s_{\text{stop}}$ values observed on the hardware. We can understand this as follows.  When the local field is stopped, the Hamiltonian is composed of a weak transverse field and the longitudinal field noise.  If the latter dominates and the state is effectively one of the two computational basis states (which it should be because the evolution up to this point has been almost adiabatic), then the remaining part of the anneal will not significantly affect the state (it precesses around the slowly time-varying axis defined by the Hamiltonian). 
However, for large values of $h$ and $\tau$ it predicts a transition at an earlier time and with a more gradual slope compared to what is observed on the hardware. Note that the coupling with the environment is expected to play a more prominent role in this latter scenario. It is worth noting that in all cases, the saturation point is reproduced accurately in agreement with the capacity of this model to fit the $h$-sweep experiment in Fig~\ref{fig:hgk_anneal}.

Finally, we observe that the open noisy model in Fig~\ref{fig:ssweep} is able to reproduce faithfully all the features that we observe from the hardware data. In particular, the transient location and slope are correctly replicated for large values of input field and large annealing time. This indicates that both the dissipative dynamics and the longitudinal field noise strongly affect the dynamics after $s_{\text{stop}}$. Specifically, the dissipative dynamics drives the system towards the thermal state of a Hamiltonian with a decaying $x$ component and fixed $z$ component.

\begin{figure*}[t]
    \centering
    \subfloat[open noiseless model]{
        \centering
        \includegraphics[width=.32\linewidth]{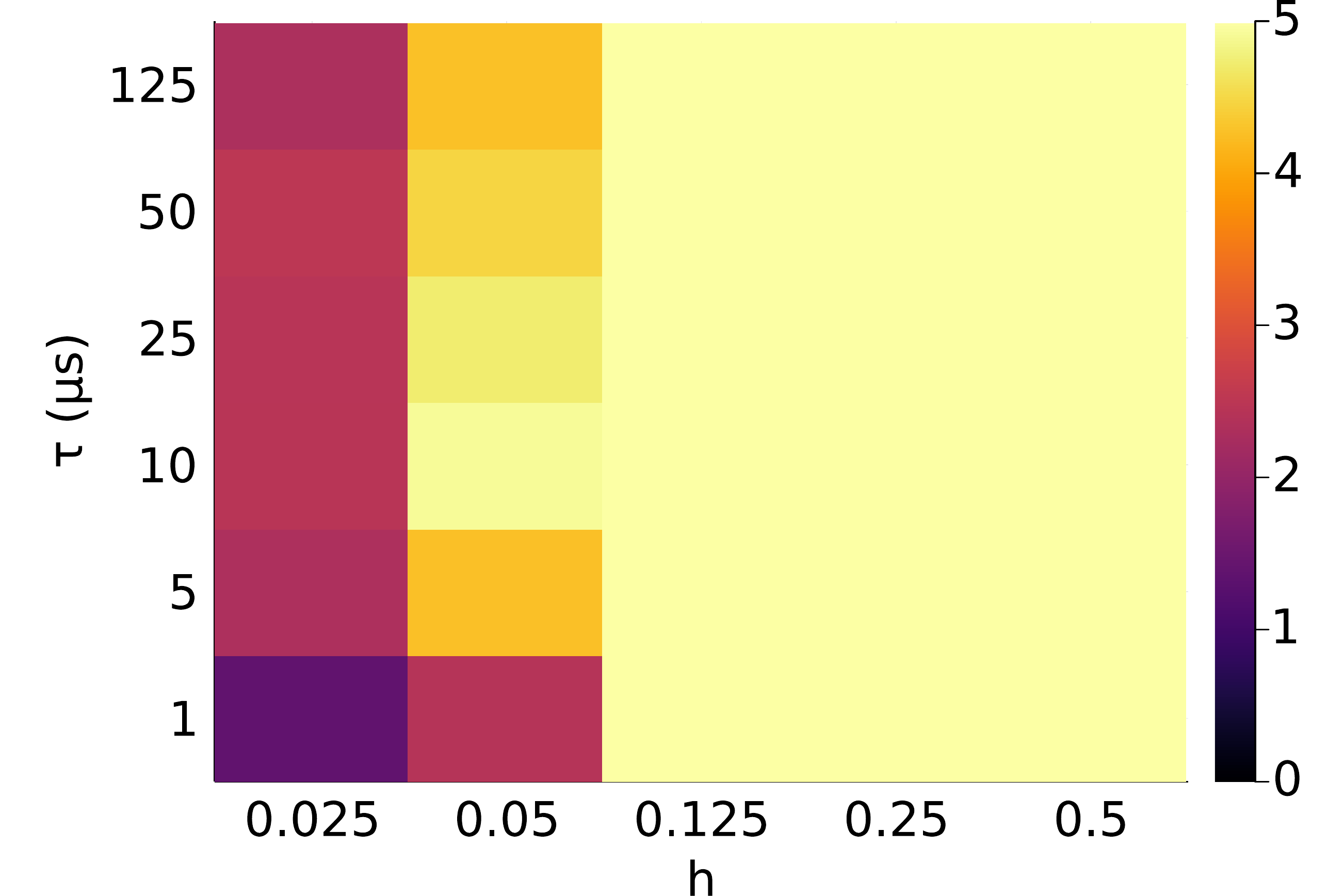}
        \label{fig:ssweep_summary_open_noisy}
    }
    \subfloat[closed noisy model]{
        \centering
        \includegraphics[width=.32\linewidth]{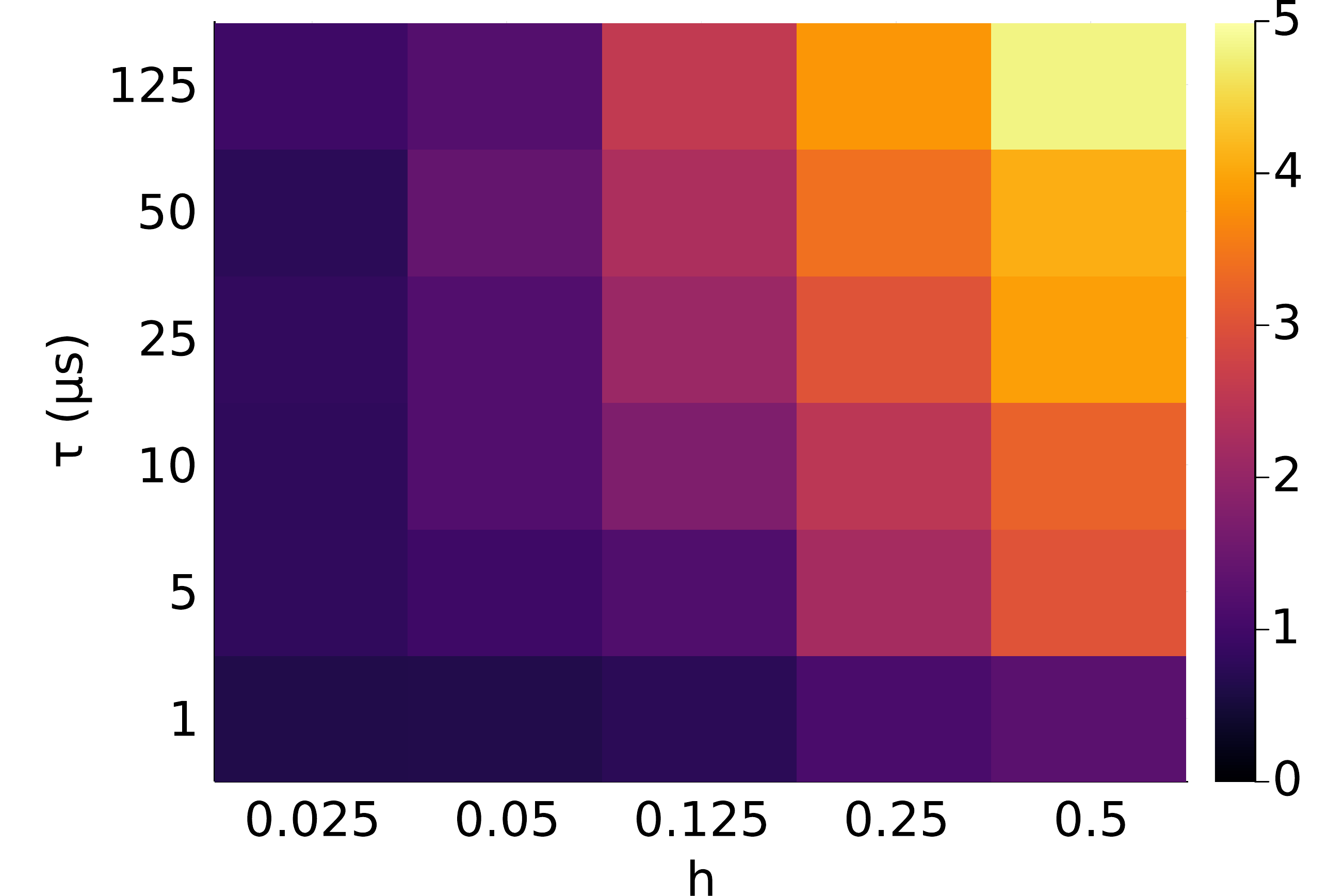}
        \label{fig:ssweep_summary_closed}
    }
    \subfloat[open noisy model]{
        \centering
        \includegraphics[width=.32\linewidth]{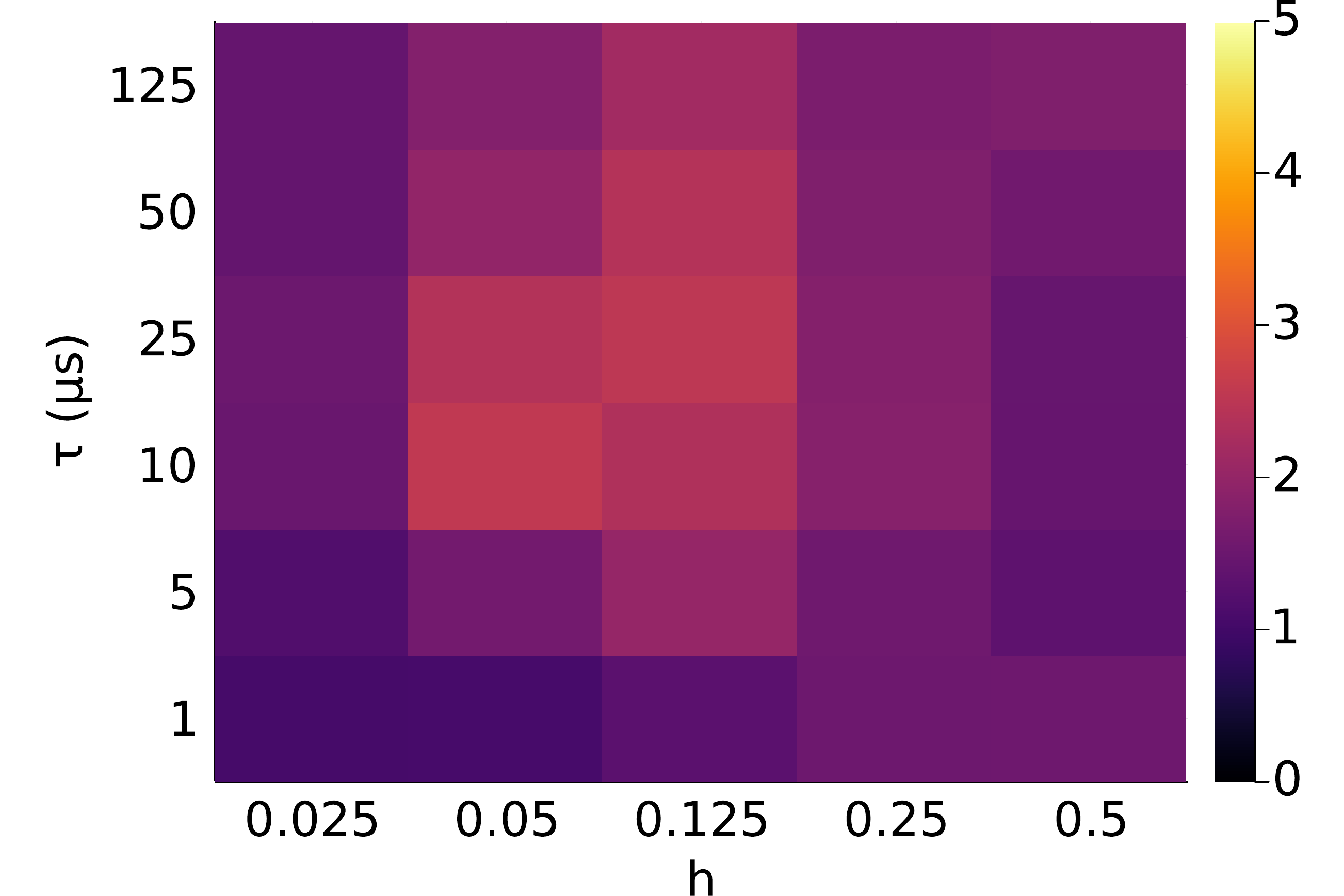}
        \label{fig:ssweep_summary_open}
    }
    \caption{These heatmaps show the $\ell_1$-distance between hardware and model measurement probabilities $P(\ket{\downarrow})$ with respect to $s_\text{stop}$ . The colorscale is identical across heatmaps.}
    \label{fig:s_sweep_summary}
\end{figure*}

The capacity of our models to reproduce experimental results for the entire range of annealing time and input magnetic fields is summarized in 
Fig~\ref{fig:s_sweep_summary}. 
The quality metric for the fit is the area between the experimental and the model measurement probabilities P(\ket{\downarrow}) with respect to $s_\text{stop}$. The previous examples from Fig~\ref{fig:ssweep} appear in the bottom-left corner, center and upper-right corner of their respective heatmaps in 
Fig~\ref{fig:s_sweep_summary}.
As seen earlier, the open noiseless model turns out to reproduce poorly the system's behavior for the entire range of parameters, while the closed noisy model tends to struggle only for the larger values of magnetic field and annealing time. In contrast, the open noisy model performs systematically better than the other two models and its fit quality is uniform over the entire range of input parameters.

At this stage, it should not be surprising that the open noisy model outperforms the other candidates. Indeed, the open noisy model has more adjustable parameters than every other model and therefore its fitting capacity is the highest among them. It remains to know if the remarkable explanation power of the open noisy model is due to overfitting or, on the contrary, if this is an indication that the model is rooted in physical ground.
We recall that the adjustable parameters are the bath coupling strength $g^2$, and the longitudinal field noise mean $\mu_{\Delta z}$ and standard deviation $\sigma_{\Delta z}$.

\begin{figure}[t]
    \centering
    \includegraphics[width=\linewidth]{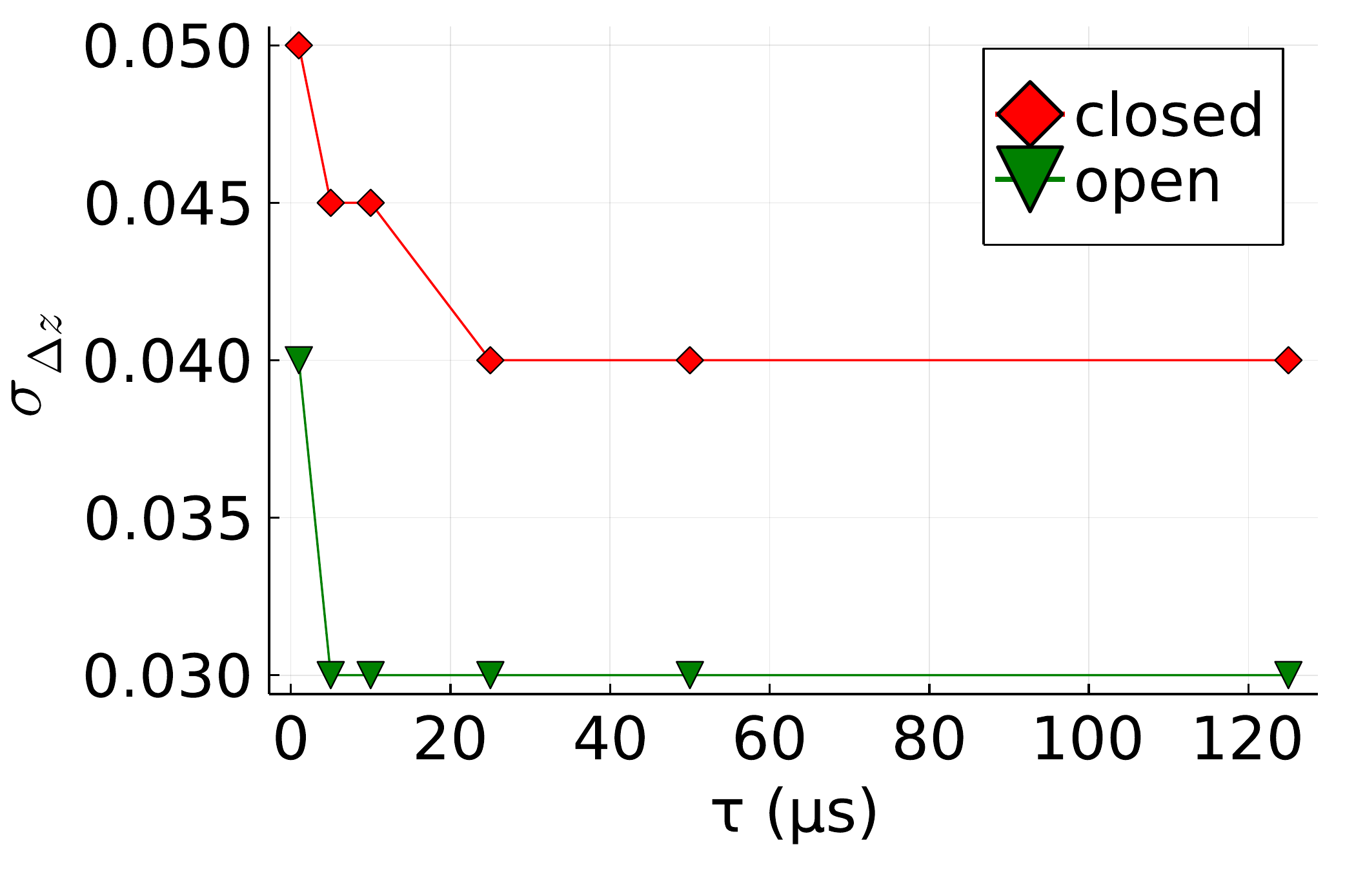}
    \caption{Fitted standard deviation $\sigma_{\Delta z}$ for the longitudinal field noise in Eq.~\eqref{eqt:fluctH} for closed-noisy and open-noisy system models. 
    }
    \label{fig:delz_values}
\end{figure}

With our fitting procedure, we find that a bath coupling strength and longitudinal field noise mean value were unaffected by the input anneal times and magnetic fields. Therefore, the bath coupling strength and longitudinal field noise mean value are set to $g^2 = 1.0 \times 10^{-6}$ and $\mu_{\Delta z}=0$ for all models. The longitudinal field noise standard deviation is found to be unaffected by the input magnetic field. Nevertheless, the standard deviation is different for the closed and open models and changing with respect to the annealing time as shown on Fig~\ref{fig:delz_values}. For the closed system, the standard deviation tends to decrease with the annealing time, a behavior that was already noticed via $h$-sweep experiments in \cite{john}. This may be caused by the inherent inability of the closed system to take into account the environment interactions that will be more prominent for longer annealing times. On the other hand, the open noisy system has a standard deviation that remains constant for annealing times bigger than $1 \mu s$. Therefore, with the exception of annealing times $\tau=1 \mu s$, the parameters of the open noisy system are set uniformly to $g^2 = 1.0 \times 10^{-6}$, $\mu_{\Delta z}=0$ and $\sigma_{\Delta z} = 0.028$.

Overly fluctuating parameters $g^2$, $\mu_{\Delta z}$ and $\sigma_{\Delta z}$ with respect to the experimental setting defined by the annealing time $\tau$ and input magnetic field $h$ would have constituted a signature of overfitting from the open noisy model. However, we see that the model parameters remain primarily constant between experiments as is expected, for they are believed to be mostly physically unaffected by the choice of experimental settings.
The different behavior of the standard deviation at $\tau=1 \mu s$ for the open noisy model is intriguing and may indicate that the longitudinal field fluctuates on time scales that are on the order of $1\mu s$. Further investigations on this question would be needed and we leave its study for future work.

\section{Conclusion}
Our experiments have shown that by implementing annealing controls with the h-gain schedule option on D-Wave's quantum annealers we obtain valuable information about the annealing dynamics. With this $h$-stop protocol, we are able to discriminate between several quantum dynamical models, highlighting that thermal fluctuations and longitudinal field noise are likely to be critical components that drive single qubits annealing dynamics. Finally, these single qubit experiments can be used in the future as efficient calibration techniques for the bath coupling strength and field noise characteristics in the simulation of more complex Hamiltonians.

{\it Acknowledgements:} We acknowledge support from the Laboratory Directed Research and Development program of Los Alamos National Laboratory under projects 20210114ER and 20210674ECR. This material is also based upon work supported by the National Science Foundation the Quantum Leap Big Idea under Grant No. OMA-1936388.

\appendix

\begin{figure*}[!t]
    \centering
    \subfloat[$L_{1}$]{
        \centering
        \includegraphics[width=.45\linewidth]{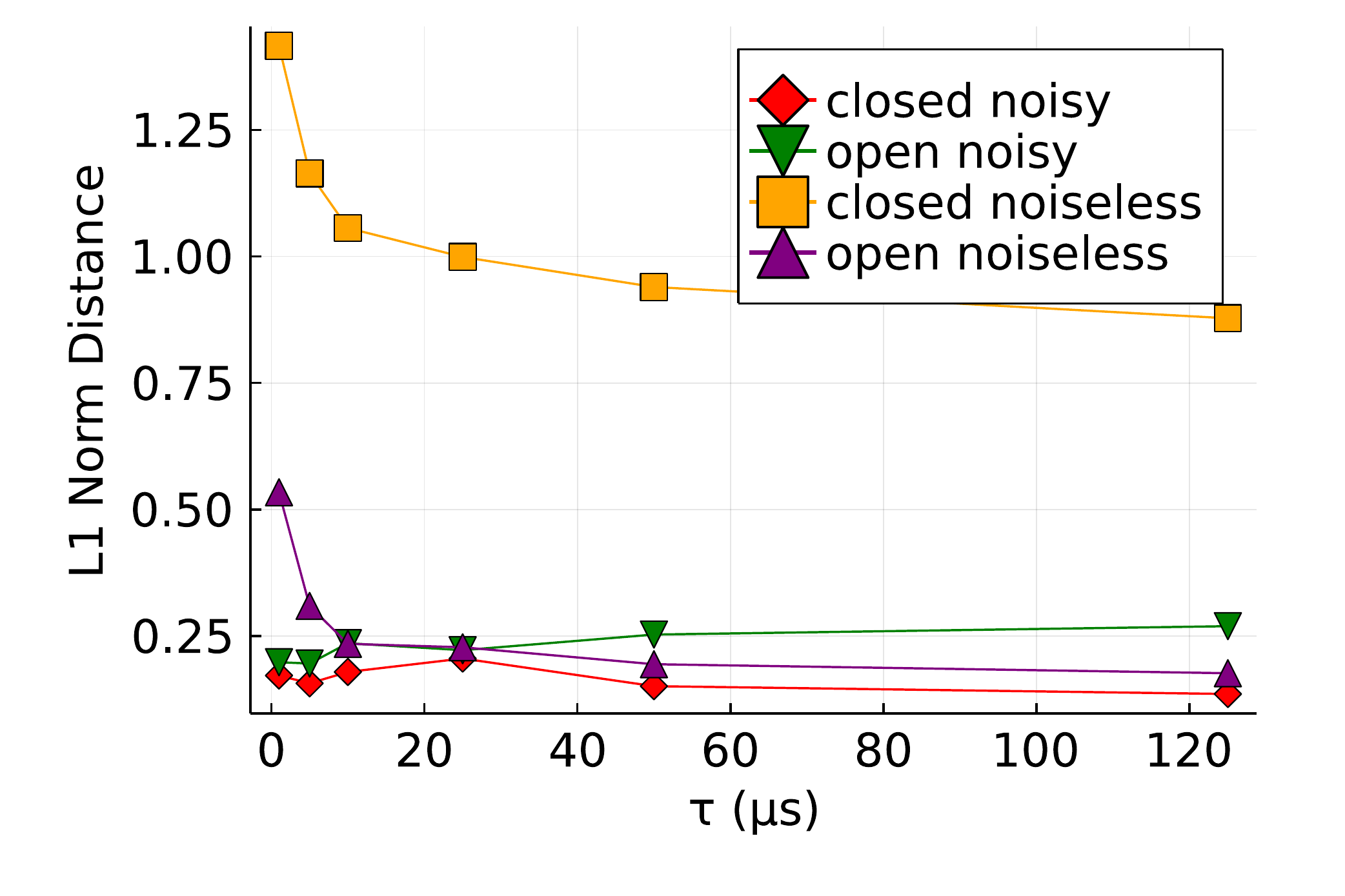}
        \label{fig:hsweep_summary_l1}
    }
    \subfloat[$L_{\infty}$]{
        \centering
        \includegraphics[width=.45\linewidth]{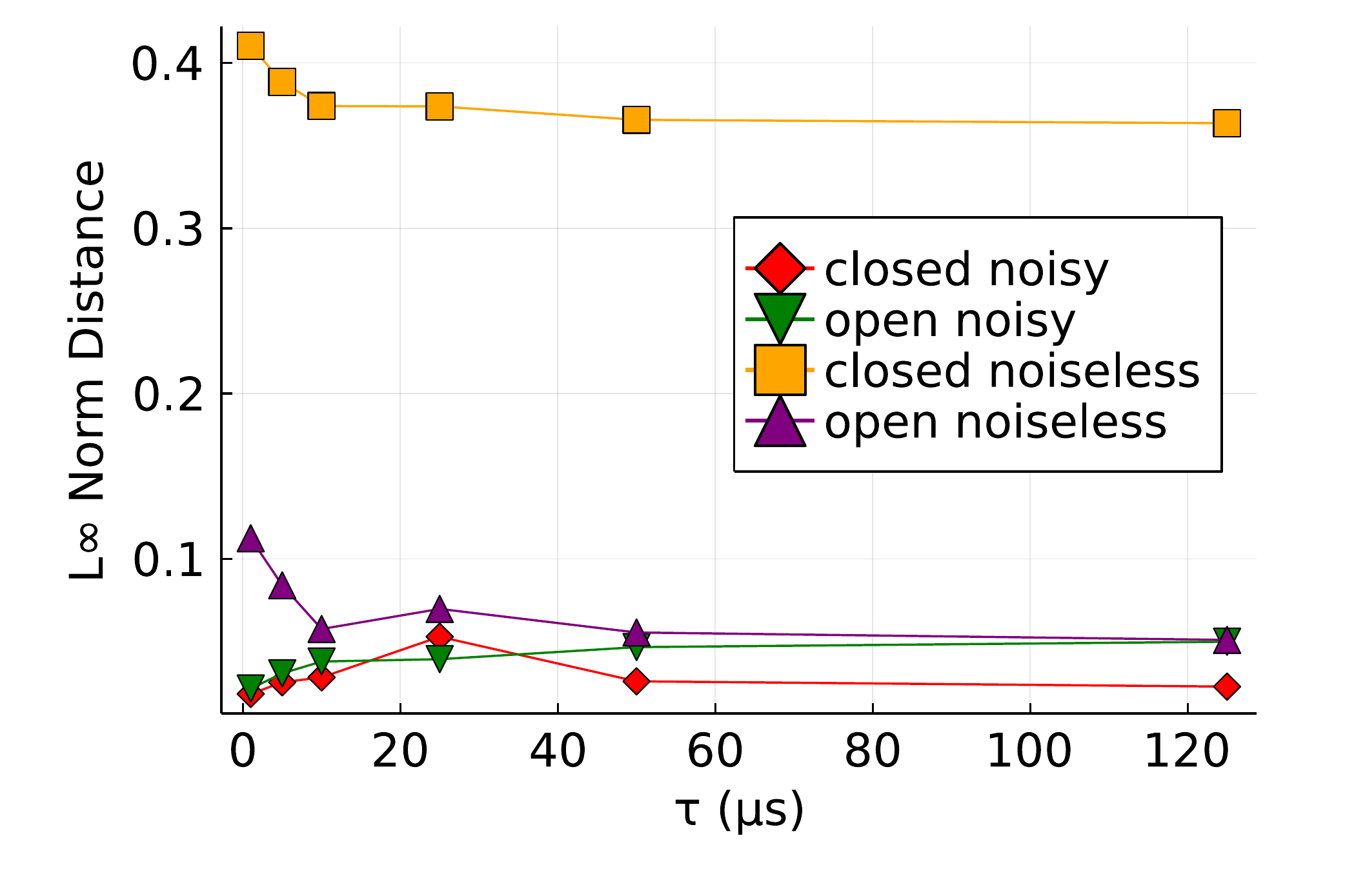}
        \label{fig:hsweep_summary_linf}
    }
    \caption{These plots show the distance from the hardware for the $h$-sweep experiment.  The parameters used to determine the models used in this experiment were also used to determine the $s$-sweep experiment, which is why the open-system seems to show a slightly worse fit.  This disparity can change depending on the distance metric and how well the $h$-stop experiment is fit.}
    \label{fig:h_sweep_summary}
\end{figure*}

\section{Model Fitting}\label{app:model_fitting}

\subsection{Parameters}
The model parameters $g^2$ and $\sigma_{\Delta z}$ were fit through grid search.  While a more refined optimization approach may have yielded a more precise fit, due to the computational overhead of performing the noise realizations required for each iteration grid search proved the most straightforward to implement.  To ensure that the grid search did not yield trends which would not make sense physically, we imposed some minor constraints on our model.  First, we required that $g^2$ be fixed for all values of $h$ and $\tau$ which were fit.  Second, we required that $\sigma_{\Delta z}$ be fixed across values of $h$, but allowed for variation between values of $t$.  This method provided a good balance that allowed the data to be fit quite well while avoiding over-fitting.  After performing the optimizations, we found that $g^2 = 1 x 10^{-6}$ fit the data well when combined with the $\sigma_{\Delta z}$ values shown in Figure \ref{fig:delz_values}.

For our distance metric, we chose to use $L_1$ distance, though other metrics such as the $L_2$ and $L_\infty$ distances could have been selected and the findings of this work would remain unchanged. The difference between the quality of the fit parameters when applied to the h-sweep model is shown in Figure \ref{fig:h_sweep_summary}.

\bibliographystyle{ieeetr}
\bibliography{ref.bib}

LA-UR-22-28691
\end{document}